\documentclass[sn-mathphys]{sn-jnl}

\jyear{2021}%

\theoremstyle{thmstyleone}%
\newtheorem{theorem}{Theorem}[section]
\newtheorem{corollary}[theorem]{Corollary}%
\newtheorem{lemma}[theorem]{Lemma}%

\theoremstyle{thmstyletwo}%

\theoremstyle{thmstylethree}%
\newtheorem{definition}[theorem]{Definition}%
\newtheorem{example}[theorem]{Example}%
\newtheorem{remark}[theorem]{Remark}%

\usepackage[utf8]{inputenc}

\usepackage{amssymb}
\usepackage{amsthm}
\usepackage{amsmath}
\usepackage{enumerate}
\usepackage{bbm}
\usepackage{graphicx}

\DeclareMathOperator*{\argmin}{arg\,min}

\begin{document}

\title[Accounting for homophily in social contact matrices]{Projecting social contact matrices to populations stratified by binary attributes with known homophily}

\author[]{\fnm{Claus} \sur{Kadelka}}\email{ckadelka@iastate.edu}

\affil[]{\orgdiv{Department of Mathematics}, \orgname{Iowa State University}, \orgaddress{\street{411 Morrill Rd}, \city{Ames}, \postcode{50011}, \state{IA}, \country{United States}}}


\abstract{Contact networks are heterogeneous. People with similar characteristics are more likely to interact, a phenomenon called assortative mixing or homophily. While age-assortativity is well-established and social contact matrices for populations stratified by age have been derived through extensive survey work, we lack empirical studies that describe contact patterns of a population stratified by other attributes such as gender, sexual orientation, ethnicity, etc. Accounting for heterogeneities with respect to these attributes can have a profound effect on the dynamics of epidemiological forecasting models. 
Here, we introduce a new methodology to expand a given e.g. age-based contact matrix to populations stratified by binary attributes with a known level of homophily. We describe a set of linear conditions any meaningful social contact matrix must satisfy and find the optimal matrix by solving a non-linear optimization problem. We show the effect homophily can have on disease dynamics and conclude by briefly describing more complicated extensions.

The available Python source code enables any modeler to account for the presence of homophily with respect to binary attributes in contact patterns, ultimately yielding more accurate predictive models.}




\maketitle

\section{Introduction}
Most social networks, e.g. physical interaction networks relevant to the spread of infectious diseases, exhibit assortative mixing, which describes the presence of more-than-expected interactions between network nodes with similar characteristics~\citep{newman2003mixing}. Assortative mixing is also known simply as assortativity, or as homophily in the case of social networks. The seminal POLYMOD study supplied empirical evidence for the presence of strong age-assortative mixing in eight European countries~\citep{mossong2008social}. Its main results are country-specific contact matrices, which describe the average number of daily contacts an individual of a certain age has with individuals of different ages, where the entire population is stratified into 5-year age bins (e.g., $0-4, 5-9, \ldots, 80+$). More recently, these contact matrices have been projected for 144 other countries including the United States, using surveys and demographic data~\citep{prem2017projecting}.

Accurate infectious disease models must account for heterogeneous contact patterns in a population. During the COVID-19 pandemic, inclusion of realistic age-mixing patterns into infectious disease models has become increasingly popular, evident in the explosion of citations of~\citep{mossong2008social,prem2017projecting}. While the available, empirically-grounded contact matrices account for homophily with respect to age, there are many other attributes (e.g., race/ethnicity, vaccine status, occupation, religion, education level or socio-economic status~\citep{mcpherson2001birds}) that are typically not included in epidemiological models. This is likely because 
there exist no empirical contact matrices that describe the assortative mixing with respect to attributes beyond age, partly because empirical studies that go beyond age present a great logistical challenge. For many of the aforementioned attributes however, we possess rough estimates of their homophily in a population. For example, it is well-established that there exists a strong level of ethnic homophily in the U.S. population~\citep{mcpherson2001birds,MollicaKellyA2003RHaI}.

In this manuscript, we describe a novel procedure that uses linear algebra and non-linear optimization techniques to infer contact matrices for populations stratified by age and additional attributes, for which estimates of their homophily exist. In this initial work, we require these attributes to be binary but extending to attributes with finitely-many values or categories should be straight-forward. In brief, we introduce a set of linear conditions a meaningful contact matrix should satisfy and show that there are typically infinitely many contact matrices that do so. To find the ``optimal" contact matrix, we therefore define an objective function and pick the contact matrix that minimizes this function. User preference can inform the choice of objective function. Throughout the entire manuscript, we follow two simple examples and show that accounting for homophily can have a strong effect on model dynamics, i.e., how a pathogen spreads throughout a population.

\section{Contact matrices}
Each individual in a population can be categorized using a multitude of attributes (e.g., age, ethnicity, education level). We can distinguish attributes by their range: some take on continuous values (e.g., age), while others are categorical (e.g., education level), discrete-valued or even binary (e.g., vaccinated against COVID-19 or not). As a whole, we can stratify a population based on a selection of attributes. In what follows, we assume all attributes take on only finitely many different values. This does not limit the usability of the developed methods, as binning can turn any continuous-valued attribute such as age into one with finitely-many choices, without losing substantial information if the number of bins is large. We begin with some basic definitions.

\begin{definition}
Let $X_1\in A_1,\ldots,X_d\in A_d$ be $d$ attributes with $2\leq \lvert A_i\rvert < \infty$. The \emph{(combined) attribute space} $A$ contains all possible $d$-tuples of attribute combinations. That is, $$A = A_1 \times \cdots \times A_d.$$
Given a population of size $N_{\text{total}}$, a (joint) \emph{distribution} (or \emph{stratification}) of that population based on the $d$ attributes is a $d$-dimensional array $N \in [0,\infty)^{A}$ with $$\sum_{i=(i_1,\ldots,i_d)\in A} N_i = N_{\text{total}}.$$
For the methods developed in this manuscript, it does not matter if $N_{\text{total}}$ describes the absolute number of individuals in the population or if $N_{\text{total}}=1$ and we consider proportions. 
\end{definition}


\begin{example}\label{ex_age}
The \emph{age distribution}, $N = (N_1,\ldots,N_m)$, of a population split into $m$ age groups is a non-negative (1-dimensional) vector of length $m$, which describes the number (or proportion) of individuals in each age group. 

For the U.S. population and the four age groups defined as 
$A = \{0-14, 15-64, 65-74, 75+\text{ years of age}\},$ the age distribution (absolute numbers in million) is
$N = (60.57, 213.61,  31.48,  22.57),$
based on 2019 U.S. Census data~\citep{bureau2019american}.
\end{example}

To describe the rates of contacts between individuals with different attribute values, we introduce the concept of a contact matrix.

\begin{definition}\label{def:contact_matrix}
Given a population stratified across the combined attribute space $A = A_1 \times \cdots \times A_d$, a \emph{contact function} (also called \emph{contact matrix} for reasons that become clear in the following remark) is a function $$C: A \times A \to [0,\infty)$$
that describes the average number of daily contacts an individual with attributes $(i_1,\ldots,i_d) \in A$ has with individuals with attributes $(j_1,\ldots,j_d) \in A$.
\end{definition}

\begin{remark}
Given a combined $d$-dimensional attribute space $A = A_1 \times \cdots \times A_d$, we can create an equivalent $1$-dimensional attribute space whose attribute values are exactly the $m:=\prod \lvert A_i\rvert$ different combinations of attribute values in $A$. This motivates the use of the term ``contact matrix" for the function $C$ defined in Defintion~\ref{def:contact_matrix}, as it can be written as an $m\times m$-matrix.
\end{remark}

\begin{remark}\label{rem_correspondence_graph}
One can think of a distribution $N$ of a population and a contact matrix $C$ as two summary statistics of an \emph{undirected} graph (i.e., a social interaction network) where the nodes (i.e., individuals) are labeled by the attribute values. 
Since the entry $C_{ij}$ in the contact matrix describes the average number of interactions (i.e., edges) connecting an individual with attribute $i\in A$ with individuals with attribute $j\in A$, the total number of interactions between individuals with attributes $i$ and $j$ is given by $N_iC_{ij}$ whenever $i\neq j$ (and $N_iC_{ii}/2$ otherwise). 
Due to the undirected nature of physical, real-world interactions, it is however also given by $N_jC_{ji}$. A meaningful contact matrix representing \emph{physical interactions}, relevant e.g. to describe the spread of an infectious disease, must therefore possess the following ``symmetry" property. 
\end{remark}

\begin{definition} (Reciprocity)
Given a combined attribute space $A$ and a corresponding distribution $N$ of a population, a contact matrix $C\in [0,\infty)^{A\times A}$ is called \emph{reciprocal} if for all $i=(i_1,\ldots,i_d),j=(j_1,\ldots,j_d) \in A$,
\begin{equation}\label{eq_symmetry}
  N_iC_{ij} = N_jC_{ji}.
\end{equation} 
This property is often also referred to as symmetry; we avoid this term due to its ambiguity in matrix theory.
\end{definition}

\begin{example}\label{ex_real1}
An age-based contact matrix has been inferred for the United States~\citep{prem2017projecting} for age groups $0-4, 5-9, \ldots, 75-79, 80+$. Using U.S. census data~\citep{bureau2019american}, this contact matrix can be condensed to the four age groups from Example~\ref{ex_age} (Table~\ref{tab:ex_real1}a). The contact matrix is not reciprocal. To see this, compare e.g. the total number of reported daily contacts between $15-64$ and $65-74$ year olds:
$$N_2C_{23} = 213.6\times 10^6\cdot 0.38 = 81.2\times  10^6 \neq 118.1\times  10^6 = N_3C_{32}.$$ 
\end{example}

\begin{table}
    \centering
    \includegraphics[width=0.9\textwidth]{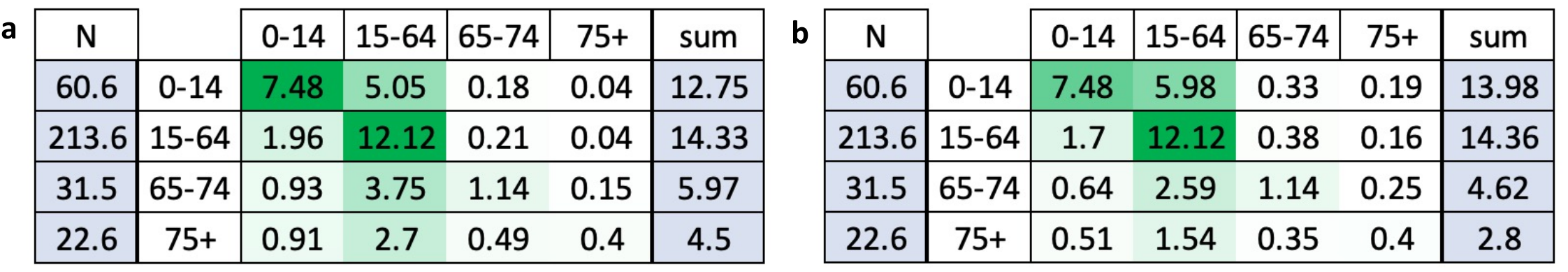}
    \caption{Age-based contact matrix for the United States, specifying the average number of daily contacts an individual of a given age group (row) has with individuals of different age groups (columns). Total population count ($N$) in millions. (a) Inferred contact matrix from~\citep{prem2017projecting}, which is not reciprocal. (b) Reciprocal contact matrix following transformation of the inferred contact matrix by Theorem~\ref{thm:reciprocal}.}
    \label{tab:ex_real1}
\end{table}

Empirical contact matrices stratified solely by age, such as those reported in the seminal POLYMOD study~\citep{mossong2008social}, are typically not reciprocal. This is because elderly people are generally more likely to report a short interaction as a contact. Extending the 1-dimensional transformation described in~\cite{funk2019combining}, an empirical contact matrix can however be easily transformed into a reciprocal one as follows. 

\begin{theorem}\label{thm:reciprocal}
Given a combined attribute space $A$, a corresponding distribution $N$ of a population, and a contact matrix $C\in [0,\infty)^{A\times A}$, the transformed contact matrix $\tilde C$ defined by
\begin{equation}\label{eq_symmetrize_funk}
   \tilde C_{ij} = \frac{N_iC_{ij} + N_jC_{ji}}{2N_i}
\end{equation}
is reciprocal.
\end{theorem}

\begin{proof}
By design of the transformation, the total number of interactions between individuals with attributes $i=(i_1,\ldots,i_d) \in A$ and $j=(j_1,\ldots,j_d) \in A$ is given by
$$N_i\tilde C_{ij}  = \frac{N_iC_{ij} + N_jC_{ji}}{2} = N_j\tilde C_{ji}.$$
\end{proof}

\begin{corollary}\label{cor:total_contacts}
The total number of contacts in a population with reciprocal contact matrix $C$ and distribution $N$ over combined attribute space $A$ is given by
$$\frac 12 \sum_{i\in A} \sum_{j\in A} N_iC_{ij}.$$
\end{corollary}

\begin{example}\label{ex_real2}
Using Theorem~\ref{thm:reciprocal}, we can transform the non-reciprocal U.S. contact matrix (Table~\ref{tab:ex_real1}a) into a reciprocal contact matrix (Table~\ref{tab:ex_real1}b).
\end{example}

\section{Homophily}

We are now in a position to (i) define the homophily of a contact matrix with respect to a binary attribute and (ii) describe the problems arising when trying to project e.g. age-specific contact matrices to populations stratified by additional binary attributes with homophily. We look first at how homophily with respect to a binary attribute can be defined for an undirected labeled graph (e.g., a physical interaction network) and use the correspondence described in Remark~\ref{rem_correspondence_graph} to define the homophily with respect to a binary attribute for a reciprocal contact matrix.

\begin{definition}\label{def:homophily_graph}
Given an undirected graph with nodes labeled by a binary attribute with \emph{prevalence} (i.e., proportion of nodes labeled \emph{true}) $p\in (0,1)$, we distinguish between two types of edges: those connecting nodes with same and with opposite attribute values. As in~\cite{kadelka2021effect}, let $\phi \in [0,1]$ denote the proportion of edges connecting nodes with same attribute values.

In the case of \emph{complete segregation}, we have $\phi = 1$ and the graph possesses two separate components. In the other extreme case of \emph{complete disassortativity}, we have $\phi = 0$ and the graph is \emph{bipartite}. Finally, in the absence of assortative mixing (i.e., \emph{no homophily}), we have $\phi = \mathbb{E}(\phi)$ where \begin{equation}\label{def:expected_phi}
\mathbb{E}(\phi) = p^2 + (1-p)^2.
\end{equation}
Extending~\cite{kadelka2021effect}, we define the \emph{homophily} $h$ with respect to the binary attribute by comparing $\phi$ to its expected value,
\begin{equation}\label{def:homophily}
h(\phi,p) = \begin{cases}
\frac{\phi-\mathbb{E}(\phi)}{1-\mathbb{E}(\phi)}\in [0,1] & \text{if }\phi\geq\mathbb{E}(\phi),\\
\frac{\phi - \mathbb{E}(\phi)}{\mathbb{E}(\phi)}\in [-1,0) & \text{if }\phi<\mathbb{E}(\phi).
\end{cases}
\end{equation}
\end{definition}

\begin{example}
If $p=2/3$ of individuals in a population are vaccinated against COVID-19, the expected number of interactions involving individuals with the same vaccination status is $\mathbb{E}(\phi) = 5/9$. Thus if $\phi = 5/9$, there exists no homophily regarding vaccination status. If $\phi = 7/9$, the population exhibits $50\%$ homophily. In the extreme case of $\phi = 1$, there exists $100\%$ homophily (i.e., complete segregation), while the other extreme case of complete disassortativity with $\phi=0$ corresponds to a homophily of $-100\%$ (or $100\%$ heterophily).
\end{example}

\begin{definition}\label{def:reduced_space}
Given a combined attribute space $A = A_1 \times \cdots \times A_d$, a corresponding distribution $N$ of a population and a contact matrix $C\in [0,\infty)^{A\times A}$, we define the \emph{reduced attribute space without the $k$th attribute} as $$A^{-k} = A_1 \times \cdots \times A_{k-1} \times A_{k+1} \times \cdots \times A_d.$$
The \emph{reduced distribution without the $k$th attribute} is a $(d-1)$-dimensional array $N^{-k} \in [0,\infty)^{A^{-k}}$ such that for all $i=(i_1,\ldots,i_{k-1},i_{k+1},\ldots,i_d) \in A^{-k}$,
$$N^{-k}_{i} = \sum_{v \in A_k} N_{i_1,\ldots,i_{k-1},v,i_{k+1},\ldots,i_d}.$$
Similarly, the \emph{reduced contact matrix without the $k$th attribute} is a function $C^{-k}: A^{-k} \times A^{-k} \to [0,\infty)$ such that for all $i,j \in A^{-k}$,
$$C^{-k}_{ij} = \sum_{v \in A_k} C_{i,(j_1,\ldots,j_{k-1},v,j_{k+1},\ldots,j_d)}.$$

If the $k$th attribute is binary, assume w.l.o.g. $A_k=\{0,1\}$ and we define the attribute's \emph{prevalence} in the population as the $(d-1)$-dimensional array $P \in [0,1]^{A^{-k}}$ such that for all $i=(i_1,\ldots,i_{k-1},i_{k+1},\ldots,i_d) \in A^{-k}$, 
$$P_{i} = \frac{N_{i_1,\ldots,i_{k-1},1,i_{k+1},\ldots,i_d}}{N^{-k}_{i}}.$$
If $d=1$, this definition corresponds exactly to the common language definition of ``prevalence", used e.g. in Definition~\ref{def:homophily_graph}.
\end{definition}


\begin{definition}\label{def:homophily_C}
Let $C\in [0,\infty)^{A\times A}$ be a reciprocal contact matrix with $A$ and $N$ as before, and let the $k$th attribute be binary. Then, the \emph{homophily of the contact matrix with respect to the binary attribute} can be defined by comparing the \emph{proportion of interactions between individuals with same attribute values}, which is given by 
\begin{equation}\label{eq_phi}
    \phi(C,N) = 
    \Bigg(\sum_{i\in A} \sum_{\substack{j \in A\\j_k = i_k}} N_iC_{ij}\Bigg) \Bigg/ \Bigg(\sum_{i\in A} \sum_{\substack{j \in A}} N_iC_{ij}\Bigg),
\end{equation}
to its expected number, $\mathbb E(\phi)$, which can be computed from the prevalence $P$ of the binary attribute, as follows.

W.l.o.g. assume that $k=d$ and $A_d =\{0,1\}$, i.e., the last attribute is binary, so that we can write $(i,v)$ to denote $(i_1,\ldots,i_{d-1},v)$ for any $i\in A^{-d}$ and $v\in A_d$. In the \emph{absence of homophily}, the binary attribute does not
affect mixing patterns. For a
given $C$, therefore, we can define a reciprocal contact matrix without homophily, denoted $C^0$, by distributing according to the prevalence $P$ the aggregated contacts in the reduced contact matrix. That is, for $i,j \in A^{-d}$ and $v\in A_d$, we have
\begin{equation}\label{eq:nohomophily}
\begin{aligned}
    C^0_{(i,v),(j,1)} &= P_j C^{-k}_{ij},\\
    C^0_{(i,v),(j,0)} &= (1-P_j) C^{-k}_{ij}.
\end{aligned}
\end{equation}

By design, we now have 
$$\mathbb{E}(\phi) = \phi(C^0,N),$$
and \emph{homophily} is defined as in Equation~\ref{def:homophily},
\begin{equation}\label{def:homophily_nd}
h(C,N) = \begin{cases}
\frac{\phi(C,N)-\mathbb{E}(\phi)}{1-\mathbb{E}(\phi)}\in [0,1] & \text{if }\phi(C,N)\geq\mathbb{E}(\phi),\\
\frac{\phi(C,N) - \mathbb{E}(\phi)}{\mathbb{E}(\phi)}\in [-1,0) & \text{if }\phi(C,N)\leq\mathbb{E}(\phi).
\end{cases}
\end{equation}
\end{definition}

\begin{remark}\label{rem:non_reciprocal_homophily}
Similar to the definition of the non-homophilic contact matrix $C^0$, we can define a contact matrix $C^h$ that has homophily $h\in (0,1]$ with respect to the $d$th binary attribute, by moving a proportion $h$ of all contacts between opposite-valued individuals to equal-valued individuals. With Equation~\ref{eq:nohomophily}, that is for $i,j \in A^{-d}$ and $v\in A_d = \{0,1\}$,

\begin{equation}\label{eq:non_reciprocal_homophily}
    \begin{aligned}
    C^h_{(i,v),(j,v)} &= C^0_{(i,v),(j,v)} + h C^0_{(i,v),(j,1-v)} = (P_{j,v} + hP_{j,1-v})C^{-d}_{ij},\\
    C^h_{(i,v),(j,1-v)} &= (1-h) C^0_{(i,v),(j,1-v)}\quad\quad\quad = (1-h)P_{j,1-v}C^{-d}_{ij},
\end{aligned}
\end{equation}
where 
\begin{equation}\label{eq_pv}
P_{j,v} = \begin{cases}P_j & \text{if $v=1$},\\
1-P_j & \text{if $v=0$}.\end{cases}
\end{equation}
While $C^h$ has homophily $h$ by definition, it is  only reciprocal in special cases. To see this, consider e.g. $(i,1),(j,1) \in A$. Reciprocity implies 
\begin{align*}
N_{(i,1)}C^h_{(i,1),(j,1)} &= N_{(j,1)}C^h_{(j,1),(i,1)}\Longleftrightarrow \\
P_iN_i^{-d} (P_j + h(1-P_j))C^{-d}_{ij} &= P_jN_j^{-d} (P_i + h(1-P_i))C^{-d}_{ji},
\end{align*}
which due to the reciprocity of $C$ is equivalent to
\begin{align*}
\qquad P_i (P_j + h(1-P_j)) &= P_j(P_i + h(1-P_i))\Longleftrightarrow \\
\qquad (P_i-P_j)h &= 0.
\end{align*}
$C^h$ is therefore only reciprocal if $h=0$ or if the prevalence is constant.
\end{remark}

\begin{example}\label{ex:running_v1}
Consider a population of size $400$, which is stratified into (i) two age groups (e.g., young and old) with age distribution $(100,300)$, and (ii) by an additional binary attribute 
with prevalence $(0.5,0.8)$ across the age groups. Further assume an age-specific contact matrix (i.e., the reduced contact matrix without the second attribute, Definition~\ref{def:reduced_space}) as in Table~\ref{fig:running_ex_v1}a. We can easily check that this contact matrix is reciprocal. Next, using the prevalence of the binary attribute, as described in Definition~\ref{def:homophily_C}, we obtain a contact matrix $C^0$ (Table~\ref{fig:running_ex_v1}b), which stratifies the population by both attributes and exhibits no homophily with respect to the binary attribute. This contact matrix is still reciprocal. As described in Remark~\ref{rem:non_reciprocal_homophily}, we can also define a contact matrix $C^h$ with any given level of homophily for the binary attribute. In the extreme case of complete segregation ($h=1$), this would result in the contact matrix shown in Figure~\ref{fig:running_ex_v1}c.

Note that in both extended contact matrices, $C^0$ and $C^h$, the total number of contacts an individual of a certain age has with individuals from each age group, and therefore also the total number of contacts (row sum), agrees with the basic age-age contact matrix, which is desirable. $C^h$ is however not reciprocal ($50\cdot 12 \neq 240\cdot 4$) because $h\neq 0$ and the prevalence of the binary attribute is not constant across age groups (Remark~\ref{rem:non_reciprocal_homophily}). Since reciprocity is a necessary condition for any \emph{physical} contact matrix, relevant e.g. to study the spread of an infectious disease, this example motivates the need for a more elaborate approach.

\begin{table}[htbp]
    \centering
    \includegraphics[width=\textwidth]{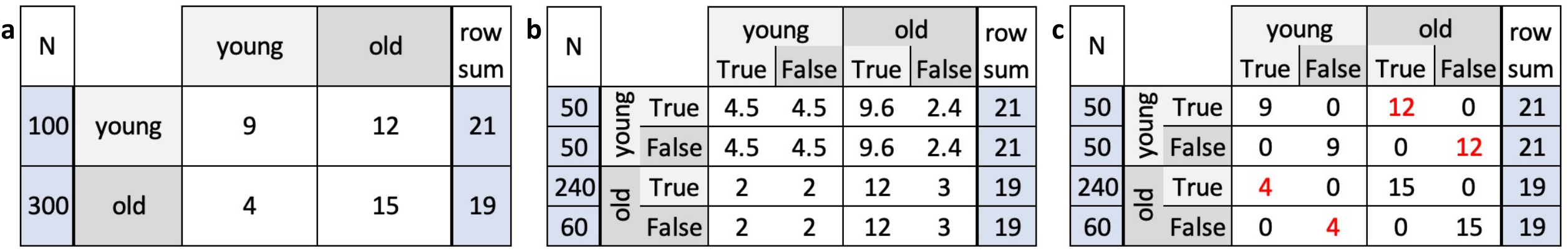}
    \caption{Contact matrices described in Example~\ref{ex:running_v1}. (a) Reciprocal contact matrix for a population stratified only by age. (b-c) Contact matrices for the same population stratified by age and an additional binary attribute. Both contact matrices possess the same number of contacts between any pair of age groups as in (a). The contact matrix in (b) exhibits no homophily and is reciprocal, while the contact matrix in (c) exhibits $100\%$ homophily but is not reciprocal (cells in red highlight where reciprocity fails). All contact matrices possess the same total number of contacts (row sum).}
    \label{fig:running_ex_v1}
\end{table}
\end{example}

\section{Extensions of contact matrices}\label{sec:4}
In this section, we will develop methods to stratify a given contact matrix by an additional binary attribute with known prevalence and homophily. We start by describing a set of necessary conditions a meaningful extended contact matrix must satisfy. Any extended contact matrix that fails to satisfy some of these conditions is not suitable to accurately describe physical contacts and can, for example, not be used to study the spread of an infectious disease.

\begin{definition}\label{prop_linear_properties}
Given a combined attribute space $A = A_1 \times \cdots \times A_d$, a corresponding distribution $N$ of a population, a reciprocal contact matrix $C\in [0,\infty)^{A\times A}$, and a binary attribute $X_{d+1}$ with known homophily $h$ and prevalence $P$ in the population, define $N^\star$ as the natural extension of the distribution (a function of $N$ and $P$) over the extended attribute space $A^\star = A \times A_{d+1}, \lvert A_{d+1}\rvert=2$ (to simplify notation, we frequently assume w.l.o.g. $A_{d+1} = \{0,1\}$). An \emph{extended contact matrix} $C^\star \in [0,\infty)^{A^\star\times A^\star}$ is \emph{meaningful} if it satisfies all of the following \emph{linear} properties:
\begin{enumerate}[(a)]
\item Reciprocity: For all $i,j \in A^\star$,
$$N^\star_i C^\star_{ij} = N^\star_j C^\star_{ji}$$
\item Same total contacts as in $C$: The total number of contacts of an individual should never depend on the value of the added binary attribute $X_{d+1}$. That is, for all $i\in A$,
$$\sum_{j\in A^\star} C^\star_{(i,0),j} = \sum_{j\in A^\star} C^\star_{(i,1),j} = \sum_{j\in A} C_{i,j}.$$
\item Same contact patterns as in $C$:  For all $i,j\in A$,
$$\sum_{v\in A_{d+1}} \frac{N^\star_{(i,v)}}{N_i} \sum_{w\in A_{d+1}} C^\star_{(i,v),(j,w)} = C_{i,j}$$
\item Homophily: For the proportion $\phi$ of contacts between individuals with same binary attribute values (Equation~\ref{eq_phi}) we have 
$$\phi = \begin{cases}
\mathbb{E}(\phi)(1-h) + h & \text{if }h\geq 0,\\
\mathbb{E}(\phi)(1+h) & \text{if }h < 0,
\end{cases}$$ where $\phi$ and $E(\phi)$ are computed as in Definition~\ref{def:homophily_C}.

If $h=1$, we require $\phi = 1$. That is, for any $i,j \in A$, $$C^\star_{(i,0),(j,1)} = C^\star_{(i,1),(j,0)} = 0.$$
Likewise, if $h=-1$, we require $\phi = 0$. That is, for any $i,j \in A$, 
$$C^\star_{(i,0),(j,0)} = C^\star_{(i,1),(j,1)} = 0.$$
\item If $N_i^\star = 0$ for some $i\in A^\star$, we can further assume $C^\star_{ij} = C^\star_{ji} = 0$ for all $j\in A^\star$. 
\end{enumerate}
\end{definition}

\begin{remark}
With $m:=\lvert A^*\rvert$, all conditions of Definition~\ref{prop_linear_properties} can be written as a linear system of $q$ equations
\begin{equation}\label{eq_linear_system}
    \mathbf X\mathbf c = \mathbf y,
\end{equation}
where $\mathbf X$ is a $q\times m^2$ matrix, $\mathbf c$ is a vector of the $m^2$ entries in the extended contact matrix $C^\star$ and $\mathbf y$ is a vector of the same dimension containing the ``right-hand"-side of the linear equations.
\end{remark}

\begin{example}\label{ex:running_v3}
Given an age-specific contact matrix $C$ with age groups young ($y$) and old ($o$) as in Table~\ref{fig:running_ex_v1}a and a binary attribute with prevalence $(0.5,0.8)$ across the age groups and homophily $h\in [0,1)$, an extended contact matrix $C^\star$ should satisfy:
\begin{enumerate}[(a)]
\item Reciprocity: \begin{align*}
    50\ C^\star_{(y,1),(y,0)} - 50\ C^\star_{(y,0),(y,1)} &=0\\
    50\ C^\star_{(y,1),(o,1)} - 240\ C^\star_{(o,1),(y,1)} &=0\\
    50\ C^\star_{(y,1),(o,0)} - 60\ C^\star_{(o,0),(y,1)} &=0\\
    50\ C^\star_{(y,0),(o,1)} - 240\ C^\star_{(o,1),(y,0)} &=0\\
    50\ C^\star_{(y,0),(o,0)} - 60\ C^\star_{(o,0),(y,0)} &=0\\
    240\ C^\star_{(o,1),(o,0)} - 60\ C^\star_{(o,0),(o,1)} &=0
\end{align*}
\item Same total contacts as in C:\begin{align*}
    C^\star_{(y,1),(y,1)} + C^\star_{(y,1),(y,0)} + C^\star_{(y,1),(o,1)} + C^\star_{(y,1),(o,0)} &=21\\
    C^\star_{(y,0),(y,1)} + C^\star_{(y,0),(y,0)} + C^\star_{(y,0),(o,1)} + C^\star_{(y,0),(o,0)} &=21\\
    C^\star_{(o,1),(y,1)} + C^\star_{(o,1),(y,0)} + C^\star_{(o,1),(o,1)} + C^\star_{(o,1),(o,0)} &=19\\
    C^\star_{(o,0),(y,1)} + C^\star_{(o,0),(y,0)} + C^\star_{(o,0),(o,1)} + C^\star_{(o,0),(o,0)} &=19
    \end{align*}
\item Same contact patterns as in C: \begin{align*}
    0.5\ C^\star_{(y,1),(y,1)} + 0.5\ C^\star_{(y,1),(y,0)} + 0.5\ C^\star_{(y,0),(y,1)} + 0.5\ C^\star_{(y,0),(y,0)} &=9\\
    0.5\ C^\star_{(y,1),(o,1)} + 0.5\ C^\star_{(y,1),(o,0)} + 0.5\ C^\star_{(y,0),(o,1)} + 0.5\ C^\star_{(y,0),(o,0)} &=12\\
    0.8\ C^\star_{(o,1),(y,1)} + 0.8\ C^\star_{(o,1),(y,0)} + 0.2\ C^\star_{(o,0),(y,1)} + 0.2\ C^\star_{(o,0),(y,0)} &=4\\
    0.8\ C^\star_{(o,1),(o,1)} + 0.8\ C^\star_{(o,1),(o,0)} + 0.2\ C^\star_{(o,0),(o,1)} + 0.2\ C^\star_{(o,0),(o,0)} &=15
    \end{align*}
\item Homophily:
\begin{align*}
50\ C^\star_{(y,1),(y,0)} + 50\ C^\star_{(y,1),(o,0)} + 50\ C^\star_{(y,0),(o,1)} + 240\ C^\star_{(o,1),(o,0)} &= (1-h) 1545,
\end{align*}
\end{enumerate}
where the number of interactions between nodes with opposite attribute values $1545 = 1-\mathbb{E}(\phi) = 1-\phi(C^0,N)$ is derived from the extended contact matrix $C^0$ without homophily shown in Table~\ref{fig:running_ex_v1}b. Thus to find an extended contact matrix that satisfies all desirable properties from Definition~\ref{prop_linear_properties}, we solve an under-determined non-homogeneous system of $15$ linear equations in $16$ variables, $\mathbf {Xc} = \mathbf y$, where $\dim \text{Nul } \mathbf X = 4$. If $h=1$, we have $C^\star_{(i,v),(j,1-v)} = 0$ for all $i,j \in \{y,o\}$ and $v \in \{0,1\}$. Here, the properties (a)-(d) yield a system of $11$ equations in $8$ variables, where $\dim \text{Nul } \mathbf X = 1$.
\end{example}

\begin{lemma}\label{lem_sol_h0}
Given $C, N, A, P$ as in Definition~\ref{prop_linear_properties}, let $C^0$ be defined as in Equation~\ref{eq:nohomophily}. That is, for $i,j \in A, v \in A_{d+1},$
\begin{equation}\label{eq:nohomophily2}
\begin{aligned}
    C^0_{(i,v),(j,1)} &= P_j C_{ij},\\
    C^0_{(i,v),(j,0)} &= (1-P_j) C_{ij}.
\end{aligned}
\end{equation}
Then, $C^0$ written as a vector $\mathbf{c^{0}} \in [0,\infty)^{m^2}$ is a non-negative solution of Equation~\ref{eq_linear_system} for homophily $h=0$.
\end{lemma}

\begin{proof}
We start by showing $C^0$ is reciprocal. Let $i,j \in A$ and $v \in A_{d+1} \cong \{0,1\}$. Remember $P_j = N^\star_{(j,1)} / N_j$ and $1-P_j = N^\star_{(j,0)} / N_j$. From the reciprocity of $C$, it follows that
\begin{align*}
    N_{(i,v)}^\star C^0_{(i,v),(j,1)} &= N_{(i,v)}^\star P_j C_{ij}\\
    &= N_{(i,v)}^\star N^\star_{(j,1)} \frac{C_{ij}}{{N_j}}\\
    &= N_{(i,v)}^\star N^\star_{(j,1)} \frac{C_{ji}}{{N_i}}\\
    &= N^\star_{(j,1)} C_{ji} \frac{N_{(i,v)}^\star}{{N_i}}\\
    &= N^\star_{(j,1)} C_{ji} \begin{cases} P_i & \text{if }v=1\\
(1-P_i) & \text{if }v=0\end{cases}\\
    &= N^\star_{(j,1)} C^0_{(j,1),(i,v)}.
\end{align*}
Equivalently, we can show $N_{(i,v)}^\star C^0_{(i,v),(j,0)} = N_{(i,v)}^\star C^0_{(j,0),(i,v)}$, which proves the reciprocity of $C^0$. Properties (b) and (c) from Definition~\ref{prop_linear_properties} follow very similarly. Finally, by design (see Definition~\ref{def:homophily_C}) $C^0$ exhibits no homophily (i.e., $\phi = \mathbb{E}(\phi)$), which means $\mathbf{c^0}$ in vector form solves  Equation~\ref{eq_linear_system} for homophily $h=0$. Non-negativity of $\mathbf{c^0}$ follows directly from $C\geq 0$ and $P_j \in [0,1]$.
\end{proof}

It is not clear if there exists a non-negative solution of Equation~\ref{eq_linear_system} for every choice of homophily $h\in [-1,1]$. It is however easy to see that the range of homophily values for which such a solution exists must be an interval.

\begin{theorem}\label{thm:sol_h_interval}
Given $C, N, A, P$ as in Definition~\ref{prop_linear_properties}, there exists $h_{\min}, h_{\max}$ with $-1\leq h_{\min} \leq 0 \leq h_{\max} \leq 1$ such that Equation~\ref{eq_linear_system} has a non-negative solution if and only if $h \in [h_{\min},h_{\max}]$.
\end{theorem}

\begin{proof}
Let $\mathbf c^0 \geq 0$ be the solution of Equation~\ref{eq_linear_system} for homophily $0$ from Lemma~\ref{lem_sol_h0}. Further, assume $\mathbf c^h$ is a non-negative solution to Equation~\ref{eq_linear_system} for homophily $h \in (0,1]$. Let $\alpha \in [0,1]$. We show that the convex combination $\mathbf c^\alpha = \alpha \mathbf c^h + (1-\alpha) \mathbf c^0$ is a non-negative solution of Equation~\ref{eq_linear_system} for homophily $\alpha h$. Let $C^0, C^h$, and $C^\alpha$ denote the contact matrices corresponding to $\mathbf c^{0}$,  $\mathbf c^{h}$, and $\mathbf c^{\alpha}$, respectively.
\begin{enumerate}[(a)]
\item Reciprocity of $C^\alpha$ follows directly from the reciprocity of $C^0$ and $C^h$: For all $i,j \in A^\star$,
$$N_i^\star C^\alpha_{ij} = \alpha N_i^\star C^h_{ij} + (1-\alpha) N_i^\star C^0_{ij} = \alpha N_j^\star C^h_{ji} + (1-\alpha) N_j^\star C^0_{ji} = N_j^\star C^\alpha_{ji}.$$
\item[(b)-(c)] Using the convex definition of $C^\alpha$, it is similarly straight-forward to show that $C^\alpha$ contains the same total contacts and same contact patterns as the original contact matrix $C$ (and as $C^0$ and $C^h$).
\item[(d)] Let $E:= \mathbb{E}(\phi)$. We know $\phi(C^0,N) = E$ and $\phi(C^h,N) = h(1-E) + E$. Due to the linear definition of $\phi$ (Equation~\ref{eq_phi}), we have 
\begin{align*}
    \phi(C^\alpha,N) &= \alpha \phi(C^h,N) + (1-\alpha) \phi(C^0,N)\\
    &= \alpha h(1-E) + \alpha E + (1-\alpha) E\\
    &= \alpha h(1-E) + E,
\end{align*}
and therefore, with Equation~\ref{def:homophily_nd}, the homophily of $C^\alpha$ is 
$$h(C^\alpha,N) = \frac{\alpha h (1-E) + E - E}{1-E} = \alpha h.$$
\end{enumerate}
Non-negativity of $C^\alpha$ follows directly from $C^h, C^0 \geq 0$. Equivalently, one can show that $\mathbf c^{\alpha}$ is a non-negative solution of Equation~\ref{eq_linear_system} for homophily $\alpha h$ given we have a solution $\mathbf c^h\geq 0$ for $h \in [-1,0)$.
\end{proof}

\begin{figure}[htbp]
    \centering
    \includegraphics[scale=0.7]{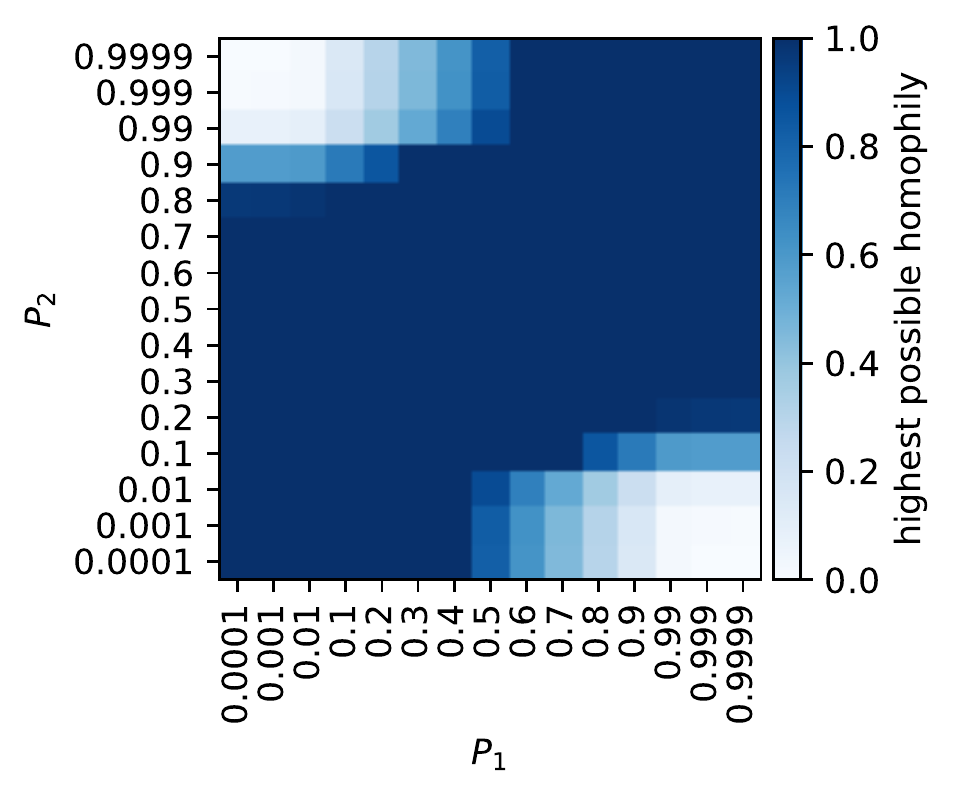}
    \caption{For different combinations of the prevalence $P$ of the added binary attribute (axes), the maximal homophily value ($h_{\max}$) that yields a non-negative solution of Equation~\ref{eq_linear_system} is shown. $C$ and $N$ are defined as in Table~\ref{fig:running_ex_v1}a.}
    \label{fig:hmax}
\end{figure}

\begin{example}
Given $C$ and $N$ as in Table~\ref{fig:running_ex_v1}a, we have $h_{\max}=1$ for most choices for the prevalence $P=(P_1,P_2)$ of the added binary attribute~(Figure~\ref{fig:hmax}). Only when $P$ is very unbalanced, do we have $h_{\max} < 1$. Moreover, $P_2$ is more important than $P_1$ in determining if there exists a non-negative solution of Equation~\ref{eq_linear_system} for homophily $h=1$. This is likely due to $N_2 > N_1$. 
\end{example}

\begin{figure}[htbp]
    \centering
    \includegraphics[scale=0.7]{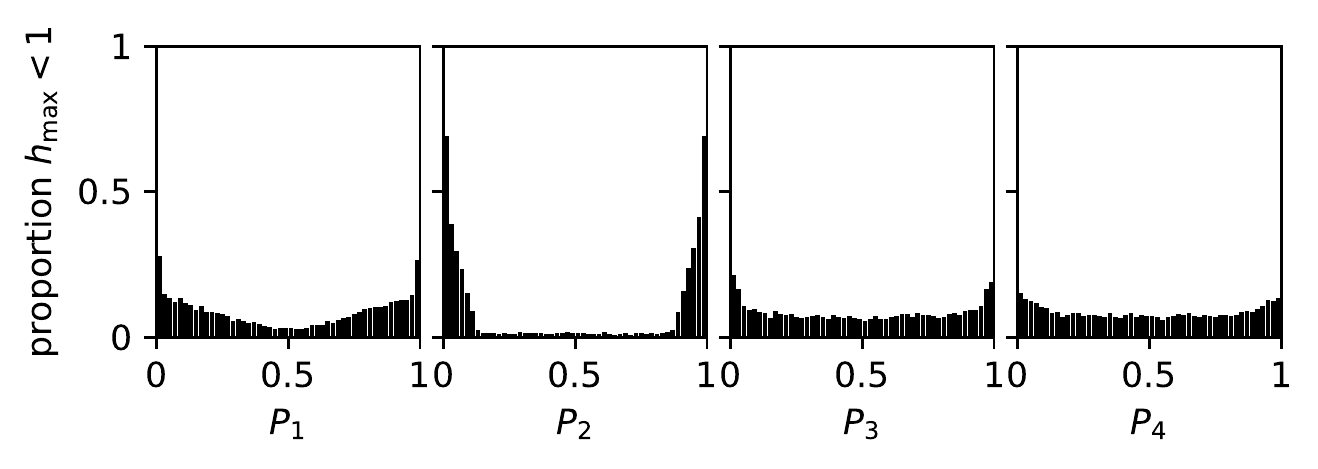}
    \caption{Proportion of random prevalence vectors $P \in [0,1]^4$ with fixed $P_i$ ($i=1,\ldots,4$ in sub panels), for which $h_{\max}<1$ for $C$ and $N$ defined for the U.S. population (see Example~\ref{ex_real3}).}
    \label{fig:hmax2}
\end{figure}

\begin{example}\label{ex_real3}
If we stratify the U.S. population with age groups $0-14, 15-64, 65-74, 75+$ (and age distribution $N$ and reciprocal contact matrix $C$ as in Table~\ref{tab:ex_real1}b) by an additional binary attribute with prevalence $P = (P_1,P_2,P_3,P_4)$, we can find a non-negative solution of Equation~\ref{eq_linear_system} for homophily $h=1$ for almost all choices of $P$. We generated $10^5$ random prevalence vectors $P \sim U(0,1)^4$ and recorded whether we were able to obtain a solution $\mathbf c^1\geq 0$. Binning by $P_i, i=1,\ldots,4$ confirms the observation made in the previous example that to ensure $h_{\max} = 1$, we primarily require a non-extreme prevalence for the age group with the largest population share, i.e., $15-64$ year olds (with prevalence $P_2$) in this example~(Figure~\ref{fig:hmax2}).
\end{example}

\begin{remark}
Given $C, N, A, P$ as in Definition~\ref{prop_linear_properties}, Theorem~\ref{thm:sol_h_interval} proves the existence of a non-negative solution $\mathbf{c}^h$ of Equation~\ref{eq_linear_system} (i.e., a contact matrix $C^h$, which satisfies all properties of Definition~\ref{prop_linear_properties}) for certain choices of homophily $h\in [-1,1]$. $C^h\geq  0$ is however normally not a unique solution. To find the ``best" contact matrix $C^\star \in [0,\infty)^{A^\star \times A^\star}$, we define an objective function $g: [0,\infty)^{A^\star \times A^\star} \to \mathbb{R}$ and solve the following optimization problem with linear equality constraints:
\begin{equation}\label{eq:optimization_over_A}
\begin{aligned}
          C^{\star} &= \argmin_{C \in [0,\infty)^{A^\star \times A^\star}} g({C})\\
          \text{subject to}\qquad \mathbf X\mathbf c&= \mathbf y,
\end{aligned}
\end{equation}
where $\mathbf c \in [0,\infty)^{m^2}$ with $m=\lvert A^*\rvert$ is, as before, the contact matrix $C$ written as a column vector. Rather than solving this problem, we can solve a related, smaller problem over the null space of $\mathbf X$ with basis $\{\mathbf{b}_1,\ldots,\mathbf{b}_r\}$, using the known solution $\mathbf{c}^h$ to formulate linear inequality constraints:
\begin{equation}\label{eq:optimization_over_null}
\begin{aligned}
          &C^{\star} = \mathbf{c}^h + \argmin_{\mathbf{x} \in \mathbb{R}^r} g\left( \mathbf{c}^h + \sum_{i=1}^r x_i\mathbf{b}_i\right)\\
          &\text{subject to}\qquad \mathbf{c}^h + \sum_{i=1}^r x_i\mathbf{b}_i \geq 0.
\end{aligned}
\end{equation}
\end{remark}

\begin{example}\label{ex:running_v4}
For $C, N, P$ defined as in Example~\ref{ex:running_v1} and homophily $h=1$, the null space of $X$ is one-dimensional. Therefore, there exist infinitely many contact matrices that satisfy all desirable properties from Definition~\ref{prop_linear_properties}. The two extreme cases are shown in Table~\ref{tab:ex_running_v4}a,b. While both contact matrices have the same age-age contact patters as in $C$, they achieve this in the most unbalanced way. For instance, to obtain an average of $C_{11}=9$ contacts for young people with young people, both matrices assign $18$ contacts to one group of the young people (those with attribute value True or False, respectively) and none to the other group. Any convex combination of these two contact matrices will be more balanced.
\end{example}

\begin{table}
    \centering
    \includegraphics[width=0.75\textwidth]{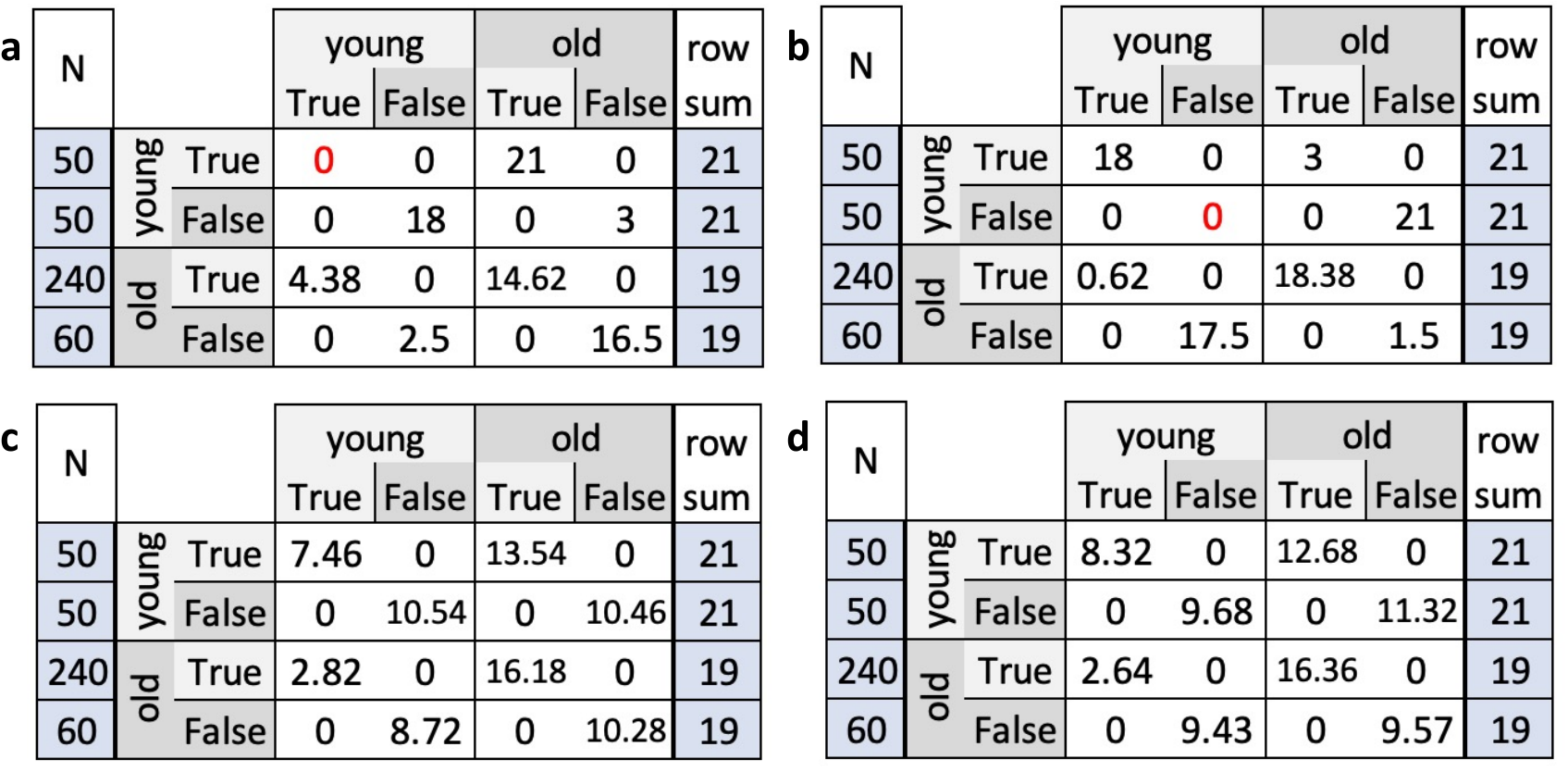}
    \caption{Selected contact matrices for $C, N, P$ defined as in Example~\ref{ex:running_v1} and homophily $h=1$. (a,b) Extreme contact matrices described in Example~\ref{ex:running_v4} where one entry describing contact of young with young people is zero (highlighted in red). (c) Contact matrix that maximizes the balance in age-age contact patterns, i.e., that minimizes $g$ defined as in Equation~\ref{eq:obj1}. (d) Least-squares solution.}
    \label{tab:ex_running_v4}
\end{table}

This example suggests an objective function that minimizes the differences in age-age contact patterns between people with opposite attribute value. Using the Euclidean distance, we can define the following objective function.

\begin{definition}\label{def:obj1}
Let $A^\star = A \times A_{d+1}$ with $ A_{d+1}=\{0,1\}$ be an extended attribute space as in Definition~\ref{prop_linear_properties}. We define an objective function $g: [0,\infty)^{A^\star \times A^\star} \to \mathbb{R}$ that measures the balance in age-age contact patterns by 
\begin{equation}\label{eq:obj1}
    g(C) = \sum_{i \in A} \sum_{j\in A}  \Big(C_{(i,1),(j,1)} + C_{(i,1),(j,0)} - C_{(i,0),(j,1)} - C_{(i,0),(j,0)}\Big)^2.
\end{equation}
\end{definition}

\begin{example}
The contact matrix shown in Table~\ref{tab:ex_running_v4}c minimizes $g$ defined as in Equation~\ref{eq:obj1}. An alternative would be to define the least-squares solution (i.e., the contact matrix $C$ that minimizes $||\mathbf X\mathbf c - \mathbf y$ from Equation~\ref{eq_linear_system}) as optimal. This would result in the contact matrix shown in Table~\ref{tab:ex_running_v4}d.
\end{example}

When perfect segregation $(h=1)$ is not desired but instead we require $h\in (0,1)$, another objective function may be more useful. Instead of only looking at the \emph{overall} homophily of a contact matrix with respect to a binary attribute, we can also define the specific homophily values for each split interaction.

\begin{definition}\label{def:obj2_split_homophily}
Let $A^\star = A \times A_{d+1}$ with $ A_{d+1}=\{0,1\}$ be an extended attribute space as in Definition~\ref{prop_linear_properties}. Given a contact matrix $C$, we define the \emph{specific homophily} of the contact patterns that an individual with attributes $(i,v)\in A^\star$ has with individuals with attributes $j\in A$ by comparing the proportion of contacts between opposite-valued individuals with the expected proportion of such contacts, given prevalence $P$. That is,
\begin{equation}\label{eq:obj2_split_homophily}
    h_{(i,v),j}(C,P) = 1 - \frac{ C_{(i,v),(j,1-v)}}{P_{j,1-v}(C_{(i,v),(j,v)} + C_{(i,v),(j,1-v)})},
\end{equation}
with $P_{j,1-v}$ defined as in Equation~\ref{eq_pv}.
\end{definition}

Ideally, the specific homophily value for each split interaction should equal the overall homophily. As outlined in Remark~\ref{rem:non_reciprocal_homophily} however, this is only possible if $h=0$ or if the prevalence is constant. Thus, we suggest as another objective to minimize the sum of the squared differences between the specific homophily values and the desired homophily.

\begin{definition}\label{def:obj2}
Let $A^\star = A \times A_{d+1}$ with $ A_{d+1}=\{0,1\}$ be an extended attribute space as in Definition~\ref{prop_linear_properties} and let $h\in [0,1]$ be the desired level of (overall) homophily with respect to the added $(d+1)$th binary attribute. We define an objective function $g_h: [0,\infty)^{A^\star \times A^\star} \to \mathbb{R}$ by setting
\begin{equation}\label{eq:obj2}
    g_h(C) = \sum_{i \in A} \sum_{j\in A} \sum_{v\in A_{d+1}} \Big(h_{(i,v),j}(C,P) - h\Big)^2,
\end{equation}
with $P$ denoting the prevalence.
\end{definition}

\begin{example}\label{ex:running_v5}
For $C, N, P$ defined as in Example~\ref{ex:running_v1} and homophily $h=0.5$, the contact matrix shown in Table~\ref{tab:ex_running_v5}a minimizes $g_h$, defined as in Equation~\ref{eq:obj2}. The specific homophily values, defined as in Equation~\ref{eq:obj2_split_homophily}, are shown in Table~\ref{tab:ex_running_v5}b. Due to the differences in the prevalence $P=(0.5,0.8)$ of the binary attribute, a contact matrix where all specific homophily values equal $h=0.5$ is not possible, as it fails the reciprocity condition.
\end{example}

\begin{table}
    \centering
    \includegraphics[width=0.75\textwidth]{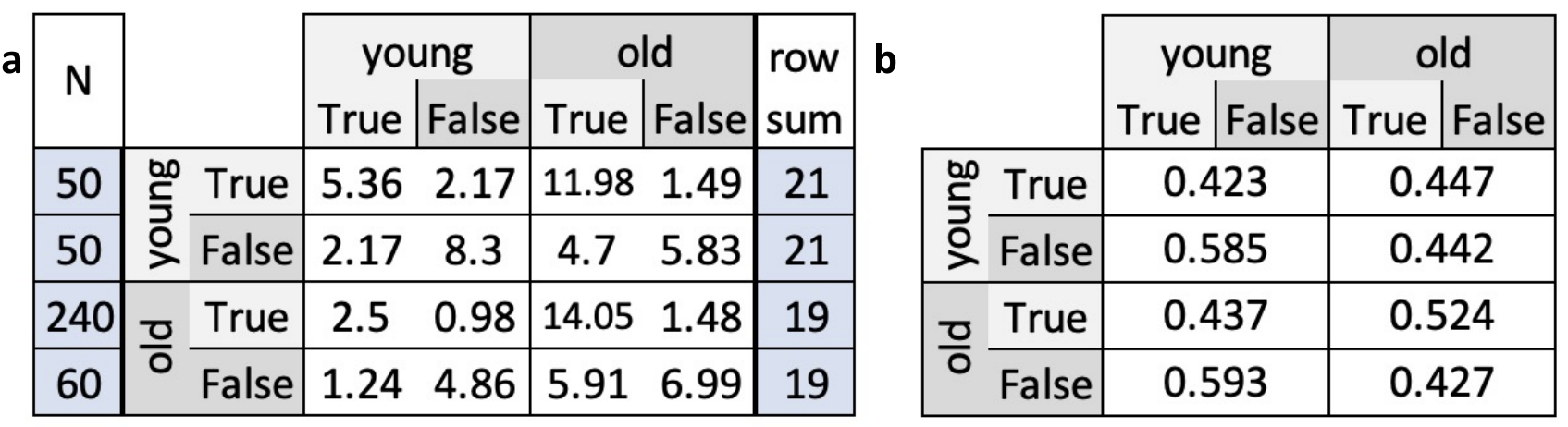}
    \caption{(a) Contact matrix that minimizes  Equation~\ref{eq:obj2}, the squared differences between specific homophily values and desired homophily $h=0.5$, for $C, N, P$ defined as in Example~\ref{ex:running_v1}. (b) Specific homophily values (Equation~\ref{eq:obj2_split_homophily}) for the contact matrix shown in (a).}
    \label{tab:ex_running_v5}
\end{table}

\section{Application}
In this section, we will compare a simple infectious disease model that only accounts for age-based mixing patterns with a more complicated model that accounts for an additional binary attribute with known homophily and prevalence. We show that disease dynamics may look quite different, especially if the homophily is high and the infection begins in one of the two groups.

\begin{definition}
A simple SIR (susceptible, infected, recovered) infectious disease model, in which the population is stratified into $m$ groups (e.g., age groups or attribute-age groups) that accounts for assortative mixing is given by 
\begin{align*}
    \frac{dS_j}{dt} &= -\lambda_j(\mathbf I)S_j\\
    \frac{dI_j}{dt} &= \lambda_j(\mathbf I)S_j - rI_j\\
    \frac{dR_j}{dt} &= rI_j
\end{align*}
for $j=1,\ldots,m$. Here, $r$ is the rate of recovery, the \emph{force of infection} takes the form 
$$\lambda_j(\mathbf I) = \lambda_j(I_1,\ldots,I_m) = \beta \sum_{k=1}^m C_{jk} I_k,$$
$\beta$ is the transmission rate and $C\in [0,\infty)^{m\times m}$ is the reciprocal contact matrix. As usual, we further assume that the population is closed with
$$\sum_{j=1}^m S_j + I_j + R_j = 1.$$
\end{definition}

\begin{figure}
    \centering
    \includegraphics[width=\textwidth]{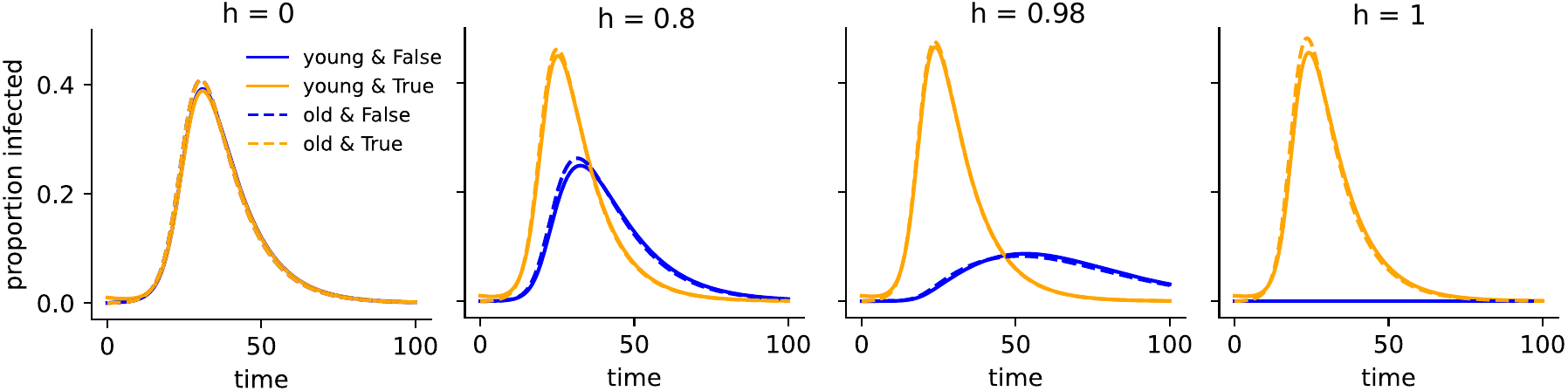}
    \caption{Effect of homophily on disease dynamics. For different levels of homophily (sub panels), the proportion of currently infected people is shown for all four sub-populations considered in Example~\ref{ex_sir}. Color distinguishes the value of the added binary attribute, while line style differentiates age.}
    \label{fig:example}
\end{figure}

\begin{example}\label{ex_sir}
To see the effect of homophily, consider $C, N, P$ as in Example~\ref{ex:running_v1} and set $\beta = 0.05, r = 0.1$. Assume there exists no pre-existing immunity in the population, $R_j(t=0) = 0$. Further, assume the disease starts in one sub-population, e.g., $$I_{\text{young \& True}}(t=0) = 0.01\ N_{\text{young \& True}},$$
while for all other sub-populations $j$ we have $I_j(0) = 0$. A comparison of the disease dynamics for different levels of homophily reveals several phenomena (Figure~\ref{fig:example}).
\begin{enumerate}
    \item Even though initially only young people are infected, age does not matter much in the dynamics, likely because the contact matrix (Table~\ref{fig:running_ex_v1}a) exhibits very low homophily with respect to age.
    \item The higher the homophily, the later and the lower the peak incidence appears among the individuals with an attribute value opposite to the one where the disease started. The presence of homophily thus introduces a delay mechanism. On the contrary, the higher the homophily, the earlier and higher the peak incidence appears among the sub-populations with the same attribute value as where the disease started. This is likely due to the requirement that the total number of contacts are constant, irrespective of the homophily. In the presence of high homophily, relatively more contacts happen between people with the same attribute value, leading to an increased spread of the disease within one group.
    \item With increasing homophily (except for the extreme case of complete segregation, $h=1$), the time until the incidence falls below a certain threshold increases. This is likely because it takes more time for the disease to manifest itself among the sub-populations with attribute values opposite to the one where the disease started.
\end{enumerate}  
\end{example}

This example clearly shows the impact high levels of homophily can have on disease dynamics and that it can be important to account for the presence of homophily.

As outlined in Remark~\ref{rem_correspondence_graph}, contact matrices can also be used to generate undirected graphs. A homophilic contact matrix thus provides a way to obtain a more realistic interaction network, on which the spread of a pathogen through a community can be studied. The details of this application are however beyond the scope of this study.

\section{More complicated extensions of contact matrices}

Thus far, we have described how to expand a given contact matrix by a binary attribute with known prevalence and homophily. In this section, we will describe one of several possible extensions where in addition to the homophilic binary attribute the population is also split into several sub-populations with differential connectivity. This is, for example, needed to accurately model different contact levels due to occupation; e.g., people with public-facing jobs on average have a lot more contacts than people with office hours. We will now describe the linear conditions a meaningful contact matrix extended by both a homophilic binary attribute and a binary attribute with differential connectivity must satisfy. 


\begin{definition}\label{prop2}
Let $A = A_1 \times \cdots \times A_d$ be a combined attribute space, $N$ a corresponding distribution of a population, $C\in [0,\infty)^{A\times A}$ a reciprocal contact matrix, and $X_{d+1}$ a binary attribute in $A_{d+1} \cong \{0,1\}$ with known homophily $h$ as in Definition~\ref{prop_linear_properties}. Let $X_{d+2}$ be a finitely-valued attribute with values $A_{d+2}$ and known differential connectivity $K \in (0,\infty)^{\lvert A_{d+2}\rvert}$. That is,  individuals with attribute value $v_2\in A_{d+2}$ have $K_{v_2}/K_{w_2}$ times the number of contacts compared to individuals with attribute value $w_2\in A_{d+2}$, assuming all other attributes in $A\times A_{d+1}$ are the same. Further, let $N^\star$ be an extended distribution of the population corresponding to the extended attribute space $A^\star = A \times A_{d+1} \times A_{d+2}$. 

An extended contact matrix $C^\star \in [0,\infty)^{A^\star\times A^\star}$ is \emph{meaningful} if it satisfies all of the following linear properties:
\begin{enumerate}[(a)]
\item Reciprocity (as in Definition~\ref{prop_linear_properties}): For all $i,j \in A^\star$,
$$N^\star_i C^\star_{ij} = N^\star_j C^\star_{ji}$$
\item Same total contacts as in $C$: 
\begin{enumerate}
    \item[(b1)] The total number of contacts of an individual should never depend on the value of the added homophilic binary attribute $X_{d+1}$. That is, for all $i\in A$ and all $v_2 \in A_{d+2}$,
$$\sum_{j\in A^\star} C^\star_{(i,0,v_2),j} = \sum_{j\in A^\star} C^\star_{(i,1,v_2),j}.$$
    \item[(b2)] Differential connectivity with respect to $X_{d+2}$ described by $K$. Fix $\bar v\in A_{d+2}$. Individuals with attribute $v_2\in A_{d+2}$ possess $K_{v_2}/K_{\bar v}$ times more contacts than individuals with attribute $\bar v\in A_{d+2}$ (and otherwise same characteristics). That is, for all $i,j\in A$ and all $v_1 \in A_{d+1}$,
$$K_{\bar v} C^\star_{(i,v_1,v_2),j} = K_{v_2} C^\star_{(i,v_1,\bar v),j}.$$
    \item[(b3)] The average total number of contacts of individuals with attribute $(i,v_1)\in A \times A_{d+1}$ is the same as in $C$. That is, for all $i\in A$, $v_1\in A_{d+1}$, 
$$\sum_{v_2\in A_{d+2}} \frac{N^\star_{(i,v_1,v_2)}}{ \sum_{w_2\in A_{d+2}} N^\star_{(i,v_1,w_2)}} \sum_{j\in A^\star} C^\star_{(i,v_1,v_2),j} = \sum_{j\in A} C_{i,j}.$$
\end{enumerate}

\item Same contact patterns as in $C$ (as in Definition~\ref{prop_linear_properties}):  For all $i,j\in A$,
$$\sum_{v_1\in A_{d+1}} \sum_{v_2\in A_{d+2}} \frac{N^\star_{(i,v_1,v_2)}}{N_i} \sum_{w_1\in A_{d+1}} \sum_{w_2\in A_{d+2}} C^\star_{(i,v_1,v_2),(j,w_1,w_2)} = C_{i,j}$$
\item Homophily (as in Definition~\ref{prop_linear_properties}): For the proportion $\phi$ of contacts between individuals with same binary attribute values (Equation~\ref{eq_phi}) we have 
$$\phi = \begin{cases}
\mathbb{E}(\phi)(1-h) + h & \text{if }h\geq 0,\\
\mathbb{E}(\phi)(1+h) & \text{if }h < 0,
\end{cases}$$ where $\phi$ and $E(\phi)$ are computed as in Definition~\ref{def:homophily_C}.

If $h=1$, we require $\phi = 1$. That is, for any $i,j \in A$ and any $v_2,w_2 \in A_{d+2}$, $$C^\star_{(i,0,v_2),(j,1,w_2)} = C^\star_{(i,1,v_2),(j,0,w_2)} = 0.$$
Likewise, if $h=-1$, we require $\phi = 0$. That is, for any $i,j \in A$ and any $v_2,w_2 \in A_{d+2}$, 
$$C^\star_{(i,0,v_2),(j,0,w_2)} = C^\star_{(i,1,v_2),(j,1,w_2)} = 0.$$
\item If $N_i^\star = 0$ for some $i\in A^\star$, we can further assume (as in Definition~\ref{prop_linear_properties}) $C^\star_{ij} = C^\star_{ji} = 0$ for all $j\in A^\star$. 
\end{enumerate}
\end{definition}

\begin{remark}
To find the ``best" meaningful extended contact matrix, one solves the same non-linear optimization problem as outlined in Section~\ref{sec:4}.
\end{remark}

\begin{remark}
Definition~\ref{prop2} also includes all linear conditions needed to find a meaningful contact matrix extended by a binary homophilic attribute, where the two attribute values have differential connectivity. Simply set $A_{d+1} = A_{d+2} \cong \{0,1\}$ and use an extended distribution $N^\star \in [0,\infty)^{A^\star}$ with $N^\star_{(i,v_1,v_2)} = 0$ whenever $v_1\neq v_2$.
\end{remark}

\section{Conclusion}
The current pandemic has revealed the importance of accurate epidemic forecasting models. One key element of these models is a contact matrix that describes the rates of mixing between different sub-populations. Age constitutes the primary attribute used to stratify populations. This is partly because age, especially for COVID-19 with its disproportionate burden on older individuals, is an important variable but also partly because most empirical work on contact matrices focuses primarily on age. This study introduces new methodology enabling modelers to expand contact matrices using binary attributes for which the prevalence and the level of homophily in a population is known or can be estimated. We showed that the disease dynamics can be very different when homophily is included in a model. A more elaborate, recent model uses this new methodology and shows that accounting for homophily with respect to ethnicity is important when designing optimal vaccine roll-out strategies~\cite{kadelka2022ethnic}.

This manuscript leaves several questions unanswered. First, we only consider binary attributes. This means any non-binary attributes need to be binarized, which may not always be feasible or desirable. An extension to non-binary attributes should be straight-forward and ideas from~\cite{feng2015elaboration} should prove useful in this effort. One possible difficulty, however, is that homophily can no longer be described by a single value, but requires $\binom m2$ variables where $m$ is the number of attribute values. Finally, we presented only a simple, simulation-based application to show the effect of homophily on disease dynamics. Related theoretical considerations, such as investigating the effect of homophily on the effective reproductive number using next-generation matrices as in~\cite{diekmann2010construction}, should provide fruitful avenues for future research.

\backmatter

\section*{Declarations}

\subsection*{Competing interests}
The author has no relevant financial or non-financial interests to disclose.

\subsection*{Code availability}
All relevant code is available at GitHub: \href{https://github.com/ckadelka/HomophilicContactMatrices}{https://github.com/ckadelka/HomophilicContactMatrices}.

\subsection*{Acknowledgments}
The author thanks Audrey McCombs for critical comments on the manuscript.

\bibliography{references}

\end{document}